\documentclass[aps,prl,reprint,superscriptaddress,letterpaperm,showpacs]{revtex4-1}
\usepackage[english]{babel}
\usepackage{amsmath}
\usepackage{amsfonts}
\usepackage{amssymb}
\usepackage{graphicx}
\usepackage{bm}
\usepackage{bbm}
\usepackage{empheq}
\usepackage{natbib}

\begin{document}

\title{Charge Relaxation in a Single Electron Si/SiGe Double Quantum Dot}
\author{K. Wang}
\author{C. Payette}
\author{Y. Dovzhenko}
\affiliation{Department of Physics, Princeton University, Princeton, New Jersey 08544, USA}
\author{P. W. Deelman}
\affiliation{HRL Laboratories LLC, 3011 Malibu Canyon Road, Malibu, California 90265, USA}
\author{J. R. Petta}
\affiliation{Department of Physics, Princeton University, Princeton, New Jersey 08544, USA}
\affiliation{Princeton Institute for the Science and Technology of Materials (PRISM), Princeton University, Princeton,
New Jersey 08544, USA}

\date{\today}

\begin{abstract}
We measure the interdot charge relaxation time ${T}_{1}$ of a single electron trapped in an accumulation mode Si/SiGe double quantum dot. The energy level structure of the charge qubit is determined using photon assisted tunneling, which reveals the presence of a low lying excited state. We systematically measure ${T}_{1}$ as a function of detuning and interdot tunnel coupling and show that it is tunable over four orders of magnitude, with a maximum of 45 $\mu$s for our device configuration.
\end{abstract}


\maketitle
Semiconductor quantum dots have been widely used as probes of fundamental quantum physics and to implement charge and spin qubits \cite{loss98,hanson07}. Coherent manipulation and two-qubit entanglement have been demonstrated in GaAs double quantum dots (DQDs) and error correction techniques such as dynamic decoupling have been employed to suppress decoherence \cite{petta05,taylor07,Bluhm11,Shulman12}. As an alternative host material, Si holds promise for ultra-coherent spin qubits due to weak spin-orbit coupling, a centrosymmetric lattice (no piezo-phonon coupling), and an established route to isotopic purification \cite{morello10, Xiao10, simmons11, Prance12, Maune12}. Spin lifetimes of 6 seconds have been measured in Si and isotopically purified $^{28}$Si crystals can support spin coherence times as long as 4 seconds \cite{morello10, Tyryshkin12}.

While Si closely approximates a ``semiconductor vacuum" for electron spins, its electronic band structure leads to potential complications that are absent in the conventional GaAs/AlGaAs two-dimensional electron gas (2DEG) system \cite{awschalom13}. First, the $\sim$ 3 times larger effective mass of electrons in Si requires depletion gate patterns to be scaled down significantly in order to achieve orbital level spacings comparable to those obtained in GaAs. Second, the band structure of bulk Si consists of six degenerate valleys, which introduces an additional decoherence pathway \cite{Ando82}. Valley degeneracy is partially lifted by uniaxial strain in a Si/SiGe heterostructure \cite{schaffler97}. However, the energy splitting between the lowest two valleys is highly sensitive to device specifics, such as step-edges in the quantum well \cite{goswami07, Friesen11, Maune12}. Detailed measurements of the low lying energy level structure, and the timescales that govern energy relaxation between these levels, are therefore needed in Si quantum dots \cite{tahan13}.

\begin{figure}[h]
\begin{center}
		\includegraphics[width=\columnwidth]{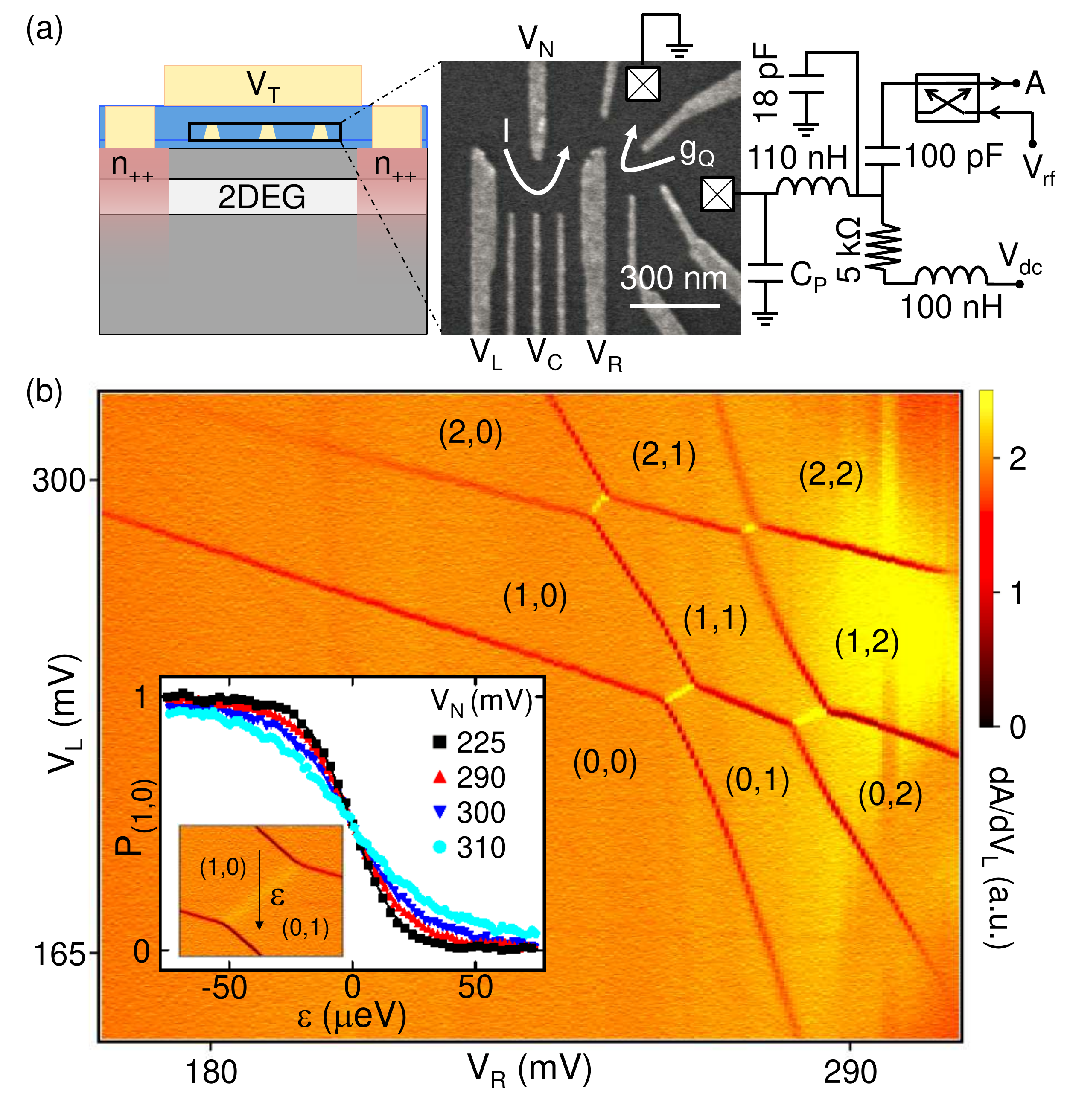}
\caption{\label{sense1} (Color online) (a) The DQD is operated by biasing a global top gate at voltage $V_{\rm T}$ to accumulate carriers in the quantum well (left). Local depletion gates define the DQD confinement potential (center). Charge sensing is performed using rf-reflectometry (right). (b) Few electron charge stability diagram visible in the derivative of the reflected rf amplitude $dA/dV_{\rm L}$. ($N_{\rm L}$,$N_{\rm R}$) indicate the number of electrons in the left and right dots. (Inset) $P_{(1,0)}$ plotted as a function of detuning, for different values of $V_{\rm N}$, showing tunable interdot tunnel coupling at the (1,0)--(0,1) interdot charge transition.}
\end{center}	
\vspace{-0.4cm}
\end{figure}

In this Letter, we systematically measure the interdot relaxation time $T_{1}$ of a single electron trapped in a Si DQD as a function of detuning $\varepsilon$ and interdot tunnel coupling $t_{\rm c}$. We demonstrate a four order of magnitude variation in $T_{1}$ using a single depletion gate and obtain $T_{1}$ = 45 $\mu$s for weak interdot tunnel couplings \cite{petta04}. We also use photon assisted tunneling (PAT) to probe the energy level structure of the single electron system, demonstrating spectroscopy with an energy resolution of $\sim$ 1 $\mu$eV. In contrast with single electron GaAs dots, we observe low lying excited states $\sim$ 55 $\mu$eV above the ground state, an energy scale that is consistent with previously measured valley splittings \cite{goswami07, Maune12}.

Measurements are performed on an accumulation mode Si/SiGe DQD. We apply a top gate voltage $V_{\rm T}$ = 2 V to accumulate carriers in a Si quantum well located $\sim$ 40 nm below the surface of the wafer [see Fig.\ 1(a)]. The resulting 2DEG has an electron density of $4\times10^{11}/{\rm cm}^{2}$ and a mobility of 70,000 ${\rm cm}^{2}/{\rm V}{\rm s}$. A 100 nm thick layer of Al$_{2}$O$_{3}$ separates the top gate from the depletion gates, which are arranged to define a DQD and a single dot charge sensor.

We first demonstrate single electron occupancy using radio frequency (rf) reflectometry \cite{schoelkopf98}. A single quantum dot is coupled to a resonant circuit with resonance frequency $f_{\rm r}$ = 431.8 MHz [Fig.\ 1(a)] and used as a high sensitivity charge detector \cite{barthel10}. The reflected amplitude $A$ is a sensitive function of the conductance of the single dot sensor, $g_{\rm Q}$, which is modulated by charge transitions in the DQD. We map out the DQD charge stability diagram in Fig.\ 1(b) by plotting the numerical derivative $dA/dV_{\rm L}$ as a function of $V_{\rm L}$ and $V_{\rm R}$. No charging transitions are observed in the lower left corner of the charge stability diagram, indicating that the DQD has been completely emptied of free electrons. We identify this charge configuration as (0,0), where ($N_{\rm L}$,$N_{\rm R}$) indicates the number of electrons in the left and right dots.

\begin{figure}[t]
\begin{center}
		\includegraphics[width=\columnwidth]{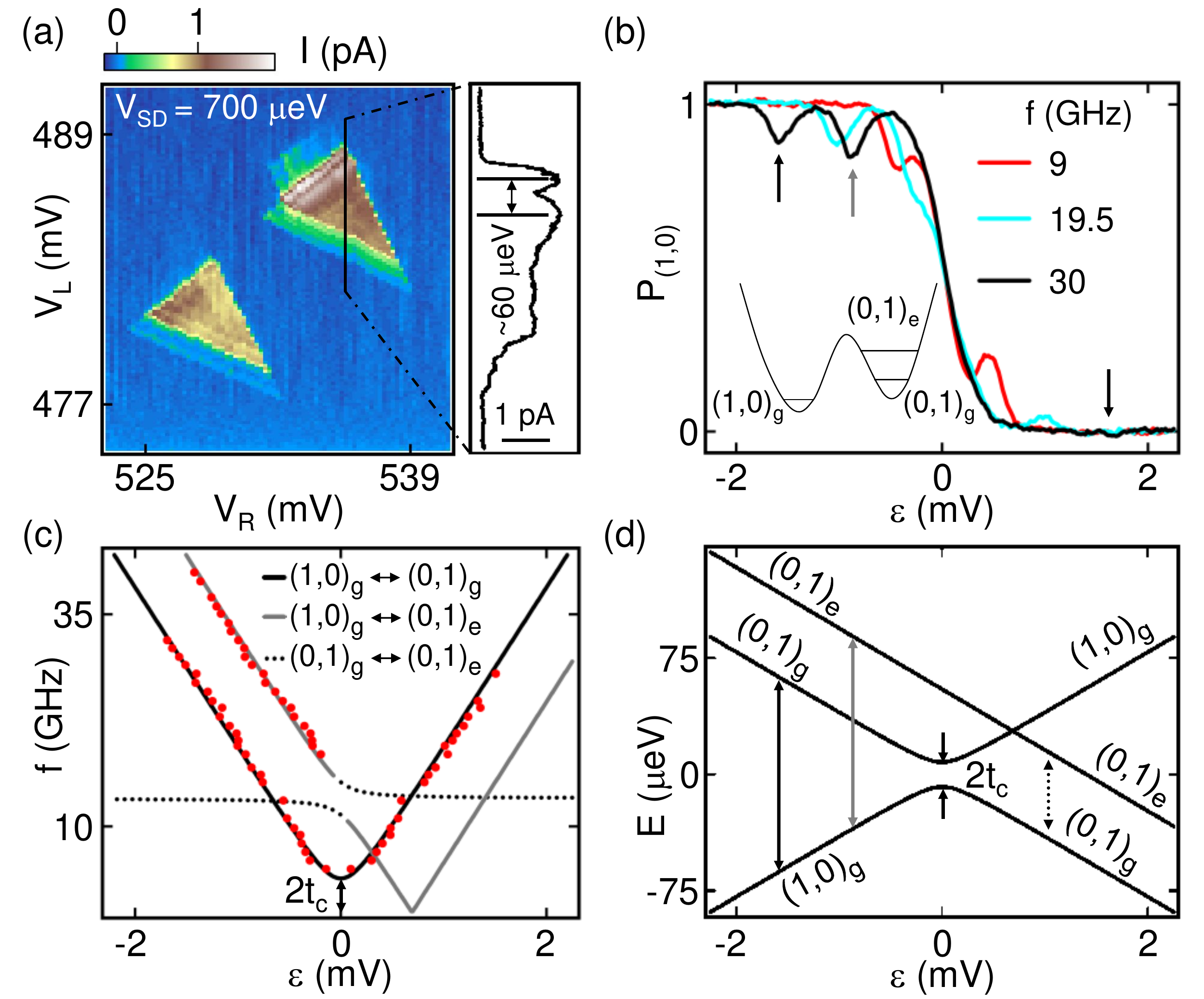}
\caption{\label{sense2} (Color online) (a) (left) Current, $I$, measured as a function of $V_{\rm L}$ and $V_{\rm R}$ near the (1,0)--(0,1) charge transition. A cut through the finite bias triangle (right) indicates the presence of a low lying excited state. (b) $P_{(1,0)}$ plotted as a function of detuning $\varepsilon$ for different excitation frequencies $f$. For $f$ $\gtrsim
$ 15 GHz, a new PAT peak emerges (grey arrow) corresponding to the $(1,0)_{\rm g}$ $\leftrightarrow$ $(0,1)_{\rm e}$ transition. The appearance of this PAT peak is accompanied by the suppression of the $(1,0)_{\rm g}$ $\leftrightarrow$ $(0,1)_{\rm g}$ PAT peak (black arrow) at positive detuning. (c) Transition frequencies as a function of detuning and (d) energy level diagram extracted from data in (c). The data in (c) are best fit with an interdot tunnel coupling $t_{\rm c}$ = 1.9 GHz and an excited state energy $\Delta$ = 55 $\mu$eV.}
\end{center}	
\end{figure}

The device is operated as a single electron charge qubit near the (1,0)--(0,1) interdot charge transition. Charge dynamics are governed by the Hamiltonian $H$ = $\frac{\varepsilon}{2}\sigma_{\rm z}+t_{\rm c}\sigma_{\rm x}$, where $\sigma_{\rm i}$ are the Pauli matrices. We demonstrate tunable interdot tunnel coupling in the single electron regime by measuring the left dot occupation $P_{(1,0)}$ as a function of detuning [Fig.\ 1(b), inset] \cite{loss98,nakamura1999,petta05}. Qubit occupation is described by
\begin{equation}
P_{(1,0)} = \frac{1}{2}\left[1-\frac{\varepsilon}{\Omega}\tanh\left(\frac{\Omega}{2k_{\rm B}T_{\rm e}}\right)\right],
\label{eq:oddreflect}
\end{equation}
where $k_{\rm B}$ is Boltzmann's constant, $T_{\rm e}$ $\sim$ 100 mK is the electron temperature, and $\Omega = \sqrt{\varepsilon^2+4t_{\rm c}^2}$ is the qubit energy splitting \cite{dicarlo04,petta04,simmons09}. With $V_{\rm N}$ = 225 mV, the interdot charge transition is thermally broadened as 2$t_{\rm c}$ $<$ $k_{\rm B}T_{\rm e}$. Increasing $t_{\rm c}$ by adjusting $V_{\rm N}$ leads to further broadening of the interdot transition. For $V_{\rm N}$ = 290, 300 and 310 mV we extract $2t_{\rm c}$ = 3.8, 5.9 and 9.0 GHz by fitting the data to Eq.\ 1. These results show that the interdot tunnel coupling can be sensitively tuned in the single electron regime in Si.

We investigate the DQD energy level structure in Fig.\ 2(a), where we plot the current $I$ as a function of $V_{\rm L}$ and $V_{\rm R}$ with a fixed source-drain bias $V_{\rm SD}$ = 700 $\mu$eV \cite{vanderwiel02}. In contrast with GaAs devices, the current in the finite bias triangles is not a smooth function of gate voltage. In particular, we observe a small resonance $\sim$ 60 $\mu$eV away from the interdot charge transition, suggesting the existence of a low-lying excited state in one of the dots. In a few electron GaAs DQD, orbital excited states are typically several meV higher in energy than the ground state \cite{Johnsonexcitedstate05}.

Higher energy resolution is obtained using PAT spectroscopy, in which microwaves drive charge transitions when the photon energy matches the qubit splitting, $h f$ = $\Omega$, where $f$ is the photon frequency and $h$ is Planck's constant. PAT transitions are directly observed as deviations from the ground state occupation in measurements of $P_{(1,0)}$ as a function of detuning [compare Fig.\ 2(b) and the inset to Fig.\ 1(b)]. For $f$ $\lesssim$ 15 GHz, the PAT peaks are symmetric around $\varepsilon$ = 0 and shift to larger detuning with increasing photon energy, consistent with a simple two level interpretation \cite{vanderWal00,petta04}. However, for $f$ $\gtrsim$ 15 GHz, an additional PAT peak emerges at negative detuning and is not accompanied by a corresponding PAT peak at positive detuning. Figure 2(c) shows the extracted transition frequencies as a function of detuning.

The data are fit using a three level Hamiltonian that includes the left dot ground state $(1,0)_{\rm g}$, the right dot ground state $(0,1)_{\rm g}$, and a right dot excited state $(0,1)_{\rm e}$, as sketched in the inset of Fig.\ 2(b) \cite{Supple13}. We obtain best fit values of $t_{\rm c}$ = 1.9 GHz and $\Delta$ = 55 $\mu$eV, consistent with the data shown in the inset of Fig.\ 1(b) and Fig.\ 2(a). Within the $\sim$ 1 $\mu$eV resolution of our measurement, we do not observe anti-crossings associated with $(0,1)_{\rm e}$. The energy eigenstates obtained from the model are plotted as a function of detuning in Fig.\ 2(d). For comparison, an excited state is observed in the left quantum dot in a second device (Device 2), with $\Delta$ = 64 $\mu$eV \cite{Supple13}. For both devices, the excited state energy is highly sensitive to $V_{\rm N}$ and $V_{\rm C}$, suggesting that it is not purely orbital in origin \cite{Friesen11}.

Several additional features observed in the data are explained by the three level model. The $(0,1)_{\rm g}$ $\leftrightarrow$ $(0,1)_{\rm e}$ intradot charge transition [dotted line, Fig.\ 2(c)] is not visible since the charge detector is only sensitive to interdot charge transitions. We also note that the $(0,1)_{\rm e}$ $\leftrightarrow$ $(1,0)_{\rm g}$ PAT peak is not visible at positive detuning. At low temperatures, the qubit population resides in the ground state $(0,1)_{\rm g}$, preventing microwave transitions from $(0,1)_{\rm e}$ to $(1,0)_{\rm g}$. Finally, the $(0,1)_{\rm g}$ $\leftrightarrow$ $(1,0)_{\rm g}$ PAT peak is suppressed when $\varepsilon$ $>$ $\Delta$ due to population trapping in $(0,1)_{\rm e}$.

We measure the interdot charge relaxation time $T_1$ by applying microwaves to $V_{\rm L}$ with a 50$\%$ duty cycle and varying the pulse period $\tau$ [Fig.\ 3(a)]. We focus on the $(1,0)_{\rm g}$ $\leftrightarrow$ $(0,1)_{\rm g}$ transition at negative detuning, where the high energy state $(0,1)_{\rm e}$ is not populated. Simulations of $P_{(1,0)}$ as a function of time, $t$, for $\tau$ = 1 $\mu$s are shown in Fig.\ 3(a) for three realistic values of $T_{1}$. During the first half of the pulse cycle, microwaves drive the $(1,0)_{\rm g}$ $\leftrightarrow$ $(0,1)_{\rm g}$ charge transition, resulting in an average $P_{(1,0)}$ = 0.5. The microwave excitation is then turned off, leading to charge relaxation during the second half of the pulse cycle, with $P_{(1,0)}$ approaching 1 on a timescale set by $T_{1}$.

\begin{figure}[t]
\begin{center}
		\includegraphics[width=\columnwidth]{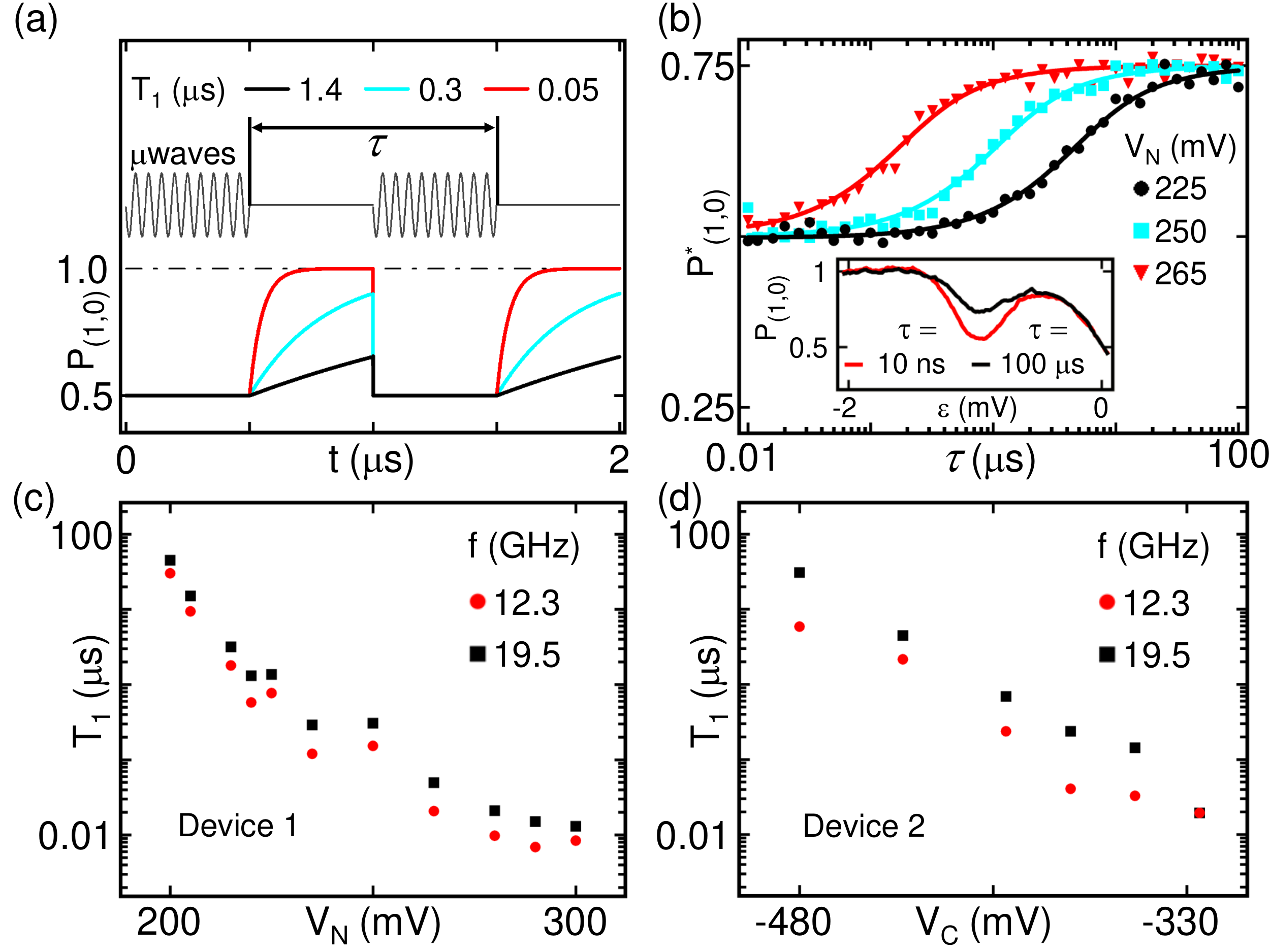}
\caption{\label{sense3} (Color online) (a) Pulse sequence used to measure $T_{1}$ and simulated qubit response. $P_{(1,0)}$ = 0.5 when resonant microwaves drive transitions between $(0,1)_{\rm g}$ and $(1,0)_{\rm g}$, and approaches 1 on a timescale set by $T_1$ when the microwaves are turned off. (b) $P^{*}_{(1,0)}$ as a function of $\tau$ extracted for different $V_{\rm N}$ at $f$ = 19.5 GHz. Fits to the data give $T_{1}$ = 1.4, 0.3 and 0.05 $\mu$s for $V_{\rm N}$ = 225, 250 and 265 mV. (Inset) Comparison of typical PAT peaks at different $\tau$, with fixed $V_{\rm N}$ = 225 mV and $f$ = 25.9 GHz. (c) $T_{1}$ as a function of $V_{\rm N}$ in Device 1. (d) $T_{1}$ as a function of $V_{\rm C}$ in Device 2.}
\end{center}	
\end{figure}

In the inset of Fig.\ 3(b), we plot $P_{(1,0)}$ as a function of detuning for $\tau$ = 10 ns and $\tau$ = 100 $\mu$s. As expected, the PAT peak is smaller for longer periods due to charge relaxation. Specifically, in the limit $\tau\ll{T}_{1}$, there is not sufficient time for relaxation to occur during the second half of the pulse cycle, leading to a time averaged value of $P_{(1,0)}$ = 0.5. In contrast, in the limit $\tau\gg T_{1}$, relaxation happens quickly, leaving $P_{(1,0)}$ = 1 for the majority of the second half of the pulse cycle. Due to experimental limitations, such as frequency dependent attenuation in the coax lines and finite pulse rise times at small $\tau$, we are unable to drive the transitions to saturation for some device configurations. To extract $T_{1}$ we therefore fit the raw $P_{(1,0)}$ data as a function of $\tau$ to the form
\begin{equation}
P_{(1,0)} = P_{\rm max}+\left(P_{\rm min}-P_{\rm max}\right)\frac{2{T}_{1}(1-e^{-\tau/(2{T}_{1})})}{\tau},
\label{eq:oddreflect}
\end{equation}
where $P_{\rm max}$ and $P_{\rm min}$ account for the limited visibility of the PAT peaks \cite{petta04,Supple13}. Extracted $T_{1}$ values are insensitive to the rescaling of the data via $P_{\rm max}$ and $P_{\rm min}$.

The interdot charge relaxation rate is strongly dependent on the interdot tunnel coupling. This variation is directly visible in the data shown in Fig.\ 3(b) for $V_{\rm N}$ = 225, 250 and 265 mV. To facilitate a direct comparison of the data, we plot the normalized electron occupation $P^{*}_{(1,0)}$ = 0.5 + 0.25 $\times$ $(P_{(1,0)} -P_{\rm min})/(P_{\rm max}-P_{\rm min})$, using the values of $P_{\rm min}$ and $P_{\rm max}$ extracted from fits to Eq.\ 2 \cite{Supple13}. In Fig.\ 3(c), we plot $T_{1}$ over a wide range of $V_{\rm N}$ for two different excitation frequencies. We see a longer characteristic relaxation time for larger interdot barrier heights, with a maximum observed value of 45 $\mu$s. The same overall trend is observed in data from Device 2 [Fig.\ 3(d)] where the interdot tunnel coupling was tuned using $V_{\rm C}$. Interdot tunnel coupling is only measurable in charge sensing when 2$t_{\rm c}$ $>$ $k_{\rm B}T_{\rm e}$ \cite{dicarlo04}. For Device 1 [see Fig.\ 3(c)] we obtain 2$t_{\rm c}$ = 2.4, 3.8 and 5.9 GHz for $V_{\rm N}$ = 280, 290 and 300 mV and for Device 2 [see Fig.\ 3(d)] we obtain 2$t_{\rm c}$ = 3.2 GHz for $V_{\rm C}$ = -325 mV.

The detuning dependence of $T_{1}$ is investigated in Fig.\ 4(a), where we plot $T_{1}$ as a function of $f$ $\propto$ $\Omega$ for the $(0,1)_{\rm g}$ $\leftrightarrow$ $(1,0)_{\rm g}$ transition \cite{nakamura1999,petta05}. Data are taken at $f$ = 12.3, 19.5, 25.9 and 30.0 GHz, as indicated by the arrows in the energy level diagram in the upper panel of Fig.\ 4(a). Our data indicate that $T_{1}$ increases weakly as a function of detuning for the range of frequencies accessible in our cryostat.

To further investigate the excited state, we measure ${T}_{1}$ for the $(0,1)_{\rm g}$ $\rightarrow$ $(1,0)_{\rm g}$ and the $(0,1)_{\rm e}$ $\rightarrow$ $(1,0)_{\rm g}$ relaxation processes at the same values of $f$ [bottom panel of Fig.\ 4(b)]. In contrast with the $(0,1)_{\rm g}$ $\rightarrow$ $(1,0)_{\rm g}$ relaxation process, $(0,1)_{\rm e}$ can relax via two distinct pathways [top panel of Fig.\ 4(b)]. The first relaxation process is a direct transition from $(0,1)_{\rm e}$ $\rightarrow$ $(1,0)_{\rm g}$ with a rate $\Gamma_{\rm e}$, while the second pathway proceeds via intradot charge relaxation to $(0,1)_{\rm g}$ with a rate $\Gamma_{\rm i}$ followed by an interdot transition to $(1,0)_{\rm g}$ with rate $\Gamma_{\rm g}^*$. We find that $(0,1)_{\rm e}$ $\rightarrow$ $(1,0)_{\rm g}$ relaxation is faster than $(0,1)_{\rm g}$ $\rightarrow$ $(1,0)_{\rm g}$ relaxation for the same energy splitting. The shorter excited state lifetime is consistent with either a fast direct relaxation rate $\Gamma_{\rm e}$ or fast intradot relaxation followed by an interdot transition. Assuming $\Gamma_{\rm g}$ = $\Gamma_{e}$ (since the level detuning is the same) and taking the measured excited state $T_{1}$ = 55 ns at $f$ = 21.0 GHz, we can make a rough lower bound estimate $\Gamma_{\rm i}$ $\sim$ $1.5\times10^{7}$ $s^{-1}$  \cite{fujisawa02}.

\begin{figure}[t]
\begin{center}
		\includegraphics[width=\columnwidth]{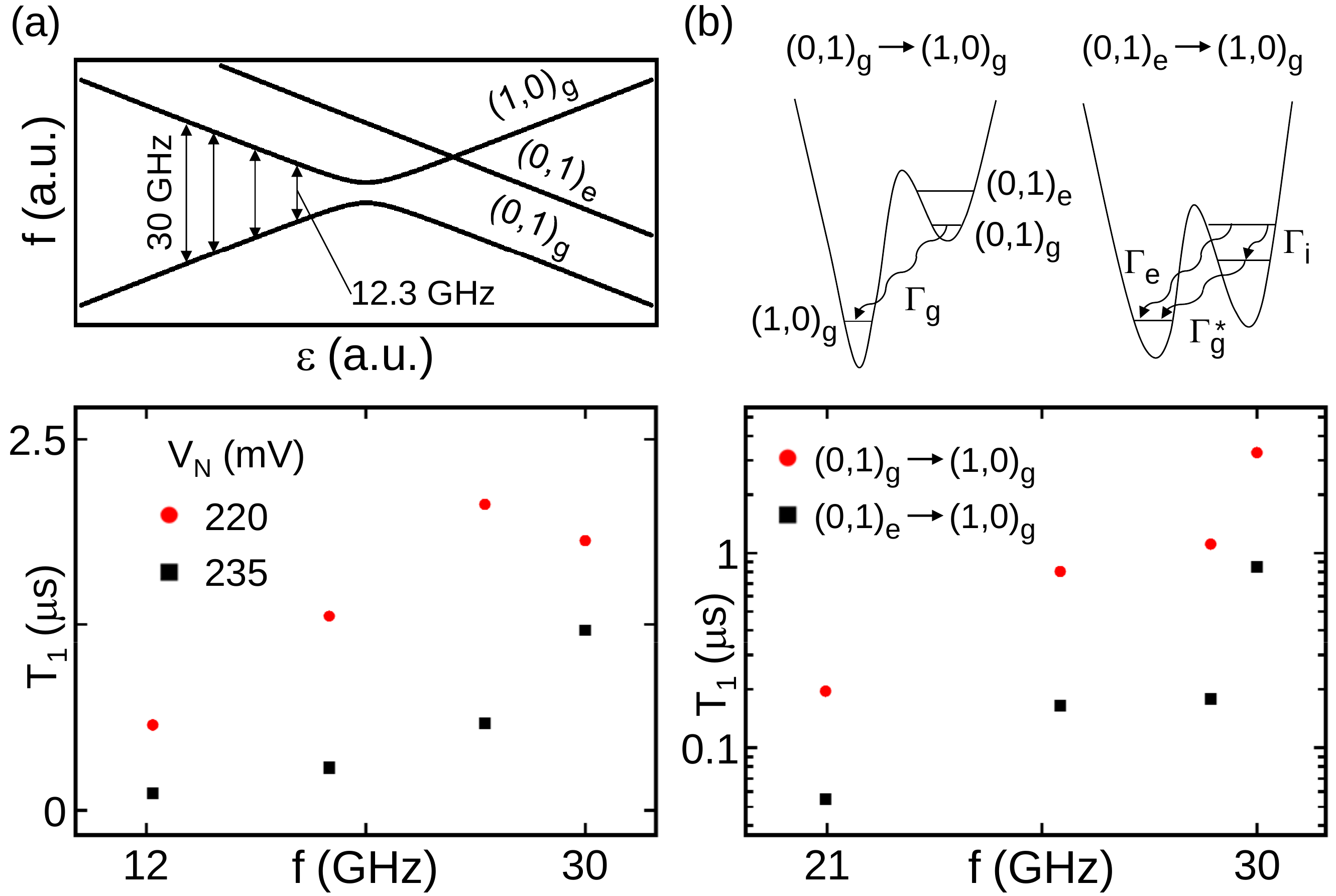}
\caption{\label{sense3} (Color online) (a) $T_{1}$ increases weakly with $f$ for the $(0,1)_{\rm g}$ $\rightarrow$ $(1,0)_{\rm g}$ transition. (b) $T_{1}$ for the $(0,1)_{\rm g}$ $\rightarrow$ $(1,0)_{\rm g}$ and $(0,1)_{\rm e}$ $\rightarrow$ $(1,0)_{\rm g}$ transitions as a function of $f$ with $V_{\rm N}$ = 250 mV (lower). There are two $(0,1)_{\rm e}$ $\rightarrow$ $(1,0)_{\rm g}$ relaxation pathways (upper).}
\end{center}	
\end{figure}

We modify the results of Raith \textit{et al.} to allow the calculation of phonon mediated charge relaxation rates considering only intravalley relaxation in the far detuned limit ($|\varepsilon| \gg t_{\rm c}$), assuming Gaussian wavefunctions for the non-hybridized charge states, with dot radius $a$ and dot separation 2$d$ \cite{Raith12,Supple13}. The electron-phonon coupling Hamiltonian in a Si quantum well takes the form
\begin{equation}
\begin{split}
H_{\rm e-ph}=&i\sum_{{}\mathbf{Q}, \lambda }^{ }\left( \frac{\hbar|\mathbf{Q}|}{2\rho Vc_{\lambda}} \right)^{\frac{1}{2}}D_{\mathbf{Q}}^{\lambda}\left(a_{\mathbf{Q},\lambda}^{\dagger}e^{i\mathbf{Q}\cdot\mathbf{r}}-a_{\mathbf{Q},\lambda}e^{-i\mathbf{Q}\cdot\mathbf{r}}\right),
\end{split}
\end{equation} where
\begin{equation}
D_{\mathbf{Q}}^{\lambda}=\left(\Xi_{\rm d}\Hat{\mathbf{e}}_{\mathbf{Q}}^{\lambda}\cdot\Hat{\mathbf{Q}}+
\Xi_{\rm u}\Hat{\rm e}_{\mathbf{Q}, \rm z}^{\lambda}\Hat{Q}_{\rm z}
\right).
\end{equation} Here $a_{\mathbf{Q},\lambda}$ ($a_{\mathbf{Q},\lambda}^{\dagger}$) are the annihilation (creation) operators for phonons belonging to branch $\lambda$ ($\lambda$ = TA1, TA2 for transverse phonons and $\lambda$ = LA for longitudinal phonons) with wave vector $\mathbf{Q}$, and speed of sound in Si $c_{\lambda}$. $V$ is the volume of the Si quantum well layer and $\rho$ is the density of Si. $\Xi _{\rm u}$ and $\Xi _{\rm d}$ are the shear and dilation deformation potential constants and $\Hat{\mathbf{e}}_{\mathbf{Q}}^{\lambda}$ and $\Hat{\mathbf{Q}}$ are the phonon unit polarization vector and the phonon unit wave vector \cite{Raith12}.
Using realistic parameters, $T_{1}$ values calculated in this model are in order of magnitude agreement with our data \cite{Supple13}. However, the predictions are exponentially sensitive to $a$ and $d$, quantities that are difficult to accurately determine. Moreover, the model predicts a relaxation rate that increases with energy splitting for the range of detunings accessed in our experiment, following the power law $\Gamma_1 = 1/T_1 \propto \Omega^3$, whereas we observe a rate that decreases weakly with increasing detuning \cite{Supple13}. This discrepancy may be due to a detuning dependent $t_{\rm c}$ or contributions from other relaxation channels, such as charge noise \cite{Astafiev04}.

In summary, we have measured charge relaxation times in a single electron Si/SiGe DQD, demonstrating a four order of magnitude variation of $T_{1}$ with gate voltage. Energy level spectroscopy indicates the presence of a low-lying excited state. From the estimated dot radius $a$ $\sim$ 38 nm, we expect orbital level spacings on the order of 1 meV, a factor of 18 larger than the value obtained from PAT spectroscopy ($\Delta$ = 55 $\mu$eV) \cite{kouwenhoven97}. This suggests that the low-lying excited state is a valley-orbit mixed state \cite{Dzurak13}.

\begin{acknowledgments}
We acknowledge helpful discussions with Jaroslav Fabian, Xuedong Hu, Debayan Mitra, Martin Raith, Peter Stano, James Sturm, and Charles Tahan. We thank Norm Jarosik for technical contributions. Research sponsored by the United States Department of Defense with partial support from the NSF through the Princeton Center for Complex Materials (DMR-0819860) and CAREER program (DMR-0846341). C. P. acknowledges partial support from FQRNT. The views and conclusions contained in this document are those of the authors and should not be interpreted as representing the official policies, either expressly or implied, of the United States Department of Defense or the U.S. Government. Approved for public release, distribution unlimited.
\end{acknowledgments}

\end{document}